\documentclass[prb,twocolumn,showpacs]{revtex4-1}
\usepackage{graphicx}
\usepackage{amsmath}
\usepackage{amssymb}

\renewcommand{\Im}{\mathrm{Im}\,}

\renewcommand{\Im}{\mathrm{Im}\,}

\DeclareMathAlphabet{\bi}{OML}{cmm}{b}{it}
\begin{document}


\title{Thermoelectric properties of an ultra-thin topological insulator}

\author{SK Firoz Islam and Tarun Kanti Ghosh}
\affiliation{Department of Physics, Indian Institute of Technology-Kanpur,
Kanpur-208 016, India}

\begin{abstract}
Thermoelectric coefficients of an ultra-thin topological 
insulator are presented here. The hybridization 
between top and bottom surface states of a topological insulator plays a
significant role. In absence of magnetic field, thermopower increases and 
thermal conductivity decreases with increase of the hybridization energy. 
In presence of magnetic field perpendicular to the ultra-thin topological insulator,
thermoelectric coefficients exhibit quantum oscillations with inverse magnetic field,  
whose frequency is strongly modified by the Zeeman energy and phase factor is
governed by the product of the Lande $g$-factor and the hybridization energy. 
In addition to the numerical results, 
the low-temperature approximate analytical results of 
the  thermoelectric coefficients
are also provided. It is also observed that for a given magnetic field 
these transport coefficients oscillate with hybridization energy, whose frequency 
depends on the Lande $g$-factor.
\end{abstract}


\date{\today}
\maketitle

\section{Introduction}
Recently a new class of material, called topological insulator,
has been paid much attention by condensed matter physicists 
\cite{kane,bernevig,konig,moore,fu,zhang}.
Topological insulator (TI) shows the conduction of electrons on the surface 
of 3D materials otherwise behaves as an insulator. It is due to the
time-reversal symmetry possessed by materials like Bi$_{2}$Se$_{3}$, Sb$_2$Te$_3$
and Bi$_2$Te$_3$\cite{zhang}. The conducting surface states of these
material show a single Dirac cone, in which spin is always locked
perpendicular to it's momentum. The angle resolved photoemission
spectroscopy \cite{hsieh1,hsieh2,xia} or scanning tunneling microscopy
\cite{roushan} has been used to realize the single Dirac cone in
TIs. In two-dimensional electron systems, under the presence of a perpendicular magnetic 
field, electron conducts along the boundary due to the circular orbits bouncing 
off the edges, leading to skipping orbits.
However, in 3D materials, even in absence of magnetic field electron conduction
takes place on the surface. Here, strong Rashba spin-orbit coupling (RSOC) plays 
the role of magnetic field. The RSOC originates from the lack of structural 
inversion symmetry of the sample \cite{rashba,rashba1}.\\

Though there have been several experimental works on 
the surface states of TIs, one of the major obstacle in studying the
transport properties of the surface is the unavoidable contribution
of the bulk. One of the best method to minimize this problem is to grow 
TIs sample in the form of ultra-thin films,
in which bulk contribution becomes relatively very small in comparison
to the surface contribution\cite{guo,lai,kehe}. 
The transition from 3D to 2D TIs lead to
several effects which have been studied for different thickness\cite{kehe,liu}.
The ultra-thin TI not only reduces the bulk contribution, but also
possesses some new phenomenon like possible excitonic superfluidity
\cite{moore1}, unique magneto-optical
response\cite{wk1,wk2,jm} and better thermoelectric
performances\cite{mong}. Moreover, the small thickness leads to the overlap of  
the wave functions between top and bottom surfaces which introduces 
a new degree of freedom hybridization\cite{yu,lu}. However, it happens to a
certain thickness of five to ten quintuple layers\cite{kehe,cao} i.e; of the order
of 10 nm.  The oscillating exponential decay of hybridization 
induced band gap with reducing thickness in Bi$_2$Te$_3$ has been also
reported theoretically\cite{yokoyama}. The formation of Landau levels have been confirmed
by several experiments\cite{cheng,jiang} in thin TIs. Moreover, several
theoretical study on low-temperature transport properties
in a series of works\cite{cao,zyuzin,wang,tahir1,tahir2} have been reported. \\

Thermoelectric properties of materials \cite{nolas} 
have always been interesting topic for providing an additional way in exploring the details
of an electronic system.
When a temperature gradient is applied across the two ends of the electronic system,
the migration of electrons from hotter to cooler side leads to the developement
of a voltage gradient across these ends.
This voltage difference per unit temperature gradient is known as longitudinal thermopower.
In addition to this temperature gradient if a perpendicular magnetic field
is applied to the system, due to Lorentz force, a transverse electric
field is also established and gives transverse thermopower. 
In conventional 2D electronic system, Landau levels induced quantum oscillation
(Shubnikov-de Hass) in thermoelectric coefficients  has been reported theoretically as well as
experimentally in a series of works\cite{prb86,prb95,topical,maximov,arindam}.
In 3D TIs, improvement of thermoelectric performance without magnetic field have been predicted
theoretically in a series of paper\cite{sinova1,sinova2,sinova3,murakami}. 
In the newly emerged relativistic-like 2D electron system-graphene, 
thermoelectric effects have been also studied\cite{yuri,wei,das,aviskar}.

In this paper, we study the effect of hybridization on the thermopower and the 
thermal conductivity of the ultra-thin TIs in absence/presence of magnetic field.
We find thermopower increases and thermal conductivity decreases with increase of the 
hybridization energy when magnetic field is absent.
In presence of perpendicular magnetic field,
thermoelectric coefficients oscillate with inverse magnetic field. 
The frequency of the quantum oscillations is strongly modified by the Zeeman energy, 
and phase factor is determined by the product of the Lande $g$-factor and the hybridization 
energy.
The analytical expressions of the thermoelectric coefficients
are also obtained. It is also shown that
these transport coefficients oscillate with frequency depend on
hybridization energy and Lande $g$-factor.

This paper has following structure. Section II briefly discusses
energy spectrum and the density of states of the ultra-thin TI in 
absence and presence of magnetic field. In section III, we have studied
how the hybridization affects the thermoelectric coefficients for
zero magnetic field. In section IV, a complete analysis of thermoelectric coefficients
in present of magnetic field is provided with numerical and analytical results.
We provide a summary and conclusion of our work in section V.

\section{ENERGY SPECTRUM AND DENSITY OF STATES }

\subsection{Zero magnetic field case}
Let us consider a surface of an ultra-thin TI in $xy$-plane 
with $L_x\times L_y$ dimension, and carriers are Dirac fermions
occupying the top and bottom surfaces of the TI. The quantum tunneling between
top and bottom surfaces gives rise to the hybridization and 
consequently the Hamiltonian 
can be written as the symmetric and anti-symmetric combination
of both surface states as \cite{yu}

\begin{equation}
H =  \left[ \begin{array}{cc}
h(k) & 0  \\
0 & h^{*}(k) \\
\end{array} \right], 
\end{equation}
with $h(k)=\Delta_h\sigma_z+v_{F}(p_y\sigma_x-p_x\sigma_y)$.
Here ${\bf p}$ is the two-dimensional momentum operator, 
$v_{F}$ is the Fermi velocity of the Dirac fermion, 
${\mbox{\boldmath $\sigma$} }=(\sigma_x,\sigma_y,\sigma_z)$ 
are the Pauli spin matrices and $\Delta_h$ is the hybridization 
matrix element between the states of the top and bottom surfaces of the TI. 
Typical value of $ \Delta_h$ varies from $(0-10^2)$ meV depending on the 
thickness of the 3D TI \cite{kehe}.
Because of the block-diagonal nature, the above Hamiltonian can be written as 
\begin{equation}
H = v_{F}(\sigma_x p_y-\tau_z\sigma_y p_x)+\Delta_h\sigma_z, 
\end{equation}
where $\tau_z=\pm$ denotes symmetric and anti-symmetric surface states, respectively.
The energy spectrum of the Dirac electron is 
given by 
\begin{equation}
E =\lambda\sqrt{(\hbar v_F k)^2+\Delta_h^2}.
\end{equation}
Here $\lambda=\pm$ stands for electron and hole bands.
The density of states is given by
\begin{equation} \label{dos1}
D_{0}(E) = \frac{2E}{\pi \hbar^2 v_{F}^2}.
\end{equation}

\subsection{Non-zero magnetic field case}
In presence of magnetic field perpendicular to the surface, 
the Hamiltonian for Dirac electron with hybridization is
\begin{equation}
H = v_F(\sigma_x\Pi_y-\tau_z\sigma_y\Pi_x)+(\tau_z\Delta_z+\Delta_h)\sigma_z, 
\end{equation}
where  ${\bf \Pi}={\bf p}+e{\bf A}$ is the two-dimensional canonical momentum operator. 
Using Landau gauge $\vec{A}=(0,Bx,0)$, exact Landau levels can be obtained 
very easily \cite{zyuzin,tahir2}. For $n=0$, there is only one energy level which is
given as 
$E_0^{\tau_z}=-(\Delta_z+\tau_z\Delta_z)$. When integer $n \geq 1$,
there are two energy bands denoted by $+$ corresponding to the electron
and $-$ corresponding to the hole with energy
\begin{equation}
E^{\tau_z}_{n,\lambda} = \lambda\sqrt{2n(\hbar\omega_c)^2+(\Delta_z+\tau_z\Delta_h)^2},
\end{equation}
where $\omega_c=v_{F}/l$ is the cyclotron frequency with $l=\sqrt{\hbar/(eB)}$
is the magnetic length, $\Delta_z=g\mu_{B}B/2 $ with $g$ is the 
Lande $g$-factor.

The corresponding eigenstates for symmetric surface state are
\begin{equation}
\Psi^{+} _{n}({\bf r}) = \frac{e^{i k_y y}}{\sqrt{L_y }}\left(
\begin{array}{r}
c_1 \phi_{n-1}(x + x_0)
\\
c_2\phi_n (x + x_0)
\end{array}
\right) \text{,}
\end{equation}
\begin{equation}
\Psi^{-} _{n}({\bf r}) = \frac{e^{i k_y y}}{\sqrt{L_y }}\left(
\begin{array}{r}
c_2 \phi_{n-1}(x + x_0)
\\
-c_1\phi_n (x + x_0)
\end{array}
\right) \text{,}
\end{equation}
where $\phi_n(x) = (1/\sqrt{\sqrt{\pi }2^n n!l}) 
e^{-x^2/2l^2} H_n(x/l) $ is the normalized harmonic 
oscillator wave function, $x_0=-k_yl^2 $,
$c_1=\cos(\theta_{\tau_z}/2)$ and $c_2=\sin(\theta_{\tau_z}/2)$
with $\theta_{\tau_z}=\tan^{-1}[\sqrt{n}\hbar\omega_c/(\Delta_z+\tau_z\Delta_h)]$.
The anti-symmetric surface state can be obtained by exchanging $n$ and $n-1$.

We have derived approximate analytical form of density of states for
$n>1$, by using the Green's function technique which is given as (see Appendix A)                                                     
\begin{eqnarray} \label{dos_B}
D_{\tau_z}(E) & \backsimeq & \frac{D_{0}(E)}{2} 
\Big[1+2\sum_{s=1}^{\infty} \exp\Big\{-s \Big(2\pi\frac{\Gamma_0 E }
{\hbar^2\omega_c^2}\Big)^2\Big\}
\nonumber\\
&\times&\cos\Big\{\pi s \Big(E^2-\Delta_{\tau_z}^2\Big)/(\hbar\omega_c)^2\Big\}\Big],
\end{eqnarray}
where $\Delta_{\tau_z}=\Delta_z+\tau_z\Delta_h$ and $\Gamma_0$ is the impurity 
induced Landau level broadening.

\section{Thermoelectric coefficients}
In this section, we shall calculate thermoelectric 
coefficients of an ultra-thin TI in zero and non-zero magnetic fields.

\subsection{Zero-magnetic field case}
In this sub-section, the effect of hybridization
on thermopower and thermal conductivity
is presented. We follow most conventional
approach at low temperature regime. The electrical current 
density ${\bf J} $ and the thermal current density ${\bf J}_{q}$ 
for Dirac electrons can be expressed under linear response regime as
\begin{equation}
{\bf J} = Q^{11} {\bf E} + Q^{12} (-{\nabla} T)
\end{equation}
and
\begin{equation}
{\bf J}^{q} = Q^{21} {\bf E} + Q^{22}(-{\nabla} T),
\end{equation}
where ${\bf E} $ is the electric field, $\nabla T$ is the temperature gradient
 and $ Q^{ij} $ ($ i,j=1,2$) are 
the phenomenological transport coefficients. The above equations describe
the response of electronic system under the combined
effects of thermal and potential gradient. 
Moreover, $Q^{ij}$ can be expressed in terms of an integral 
$I^{(r)}$: $Q^{11}=I^{(0)}, Q^{21} = TQ^{12} = -I^{(1)}/e$, 
$Q^{22}=I^{(2)}/(e^2T)$ with
\begin{equation}
I^{(r)} = \int dE  \Big[-\frac{\partial f(E)}{\partial E}\Big]
(E-\eta)^{(r)}  \sigma(E),
\end{equation}
where $r=0,1,2$ and $f(E)=1/[1+\exp(E-\eta)\beta]$ is the 
Fermi-Dirac distribution function with $\eta$ is the chemical 
potential and  $\beta=(k_{_B}T)^{-1}$. Here, 
$\sigma(E)$ is the energy-dependent electrical conductivity.
When circuit is open i.e; for $J=0$, thermopower can be defined as $S=Q^{12}/Q^{11}$.
By using Sommerfeld expansion at low temperature regime, diffusive thermopower $S$  and  
thermal conductivity $\kappa$ can be obtained 
from Mott's relation and the Wiedemann-Franz law as
\begin{equation} \label{tp}
S = - L_0 e T \Big[ \frac{d}{dE}\ln\sigma(E) \Big]_{_{E=E_F}}
\end{equation}
and
\begin{equation}\label{ther}
\kappa = L_0 T \sigma(E_{_F}).
\end{equation}
Here, $ L_0 = (\pi^2 k_{_B}^2)/(3e^2) =2.44\times10^{-8}$ W$\Omega$K$^{-2}$ 
is the Lorentz number and $\sigma(E_{_F})$  is the electrical conductivity at the Fermi energy.

Classical Boltzmann transport equation can be used to calculate zero magnetic field
electrical conductivity, which is given by\cite{mermin}
\begin{equation}\label{boltz}
\sigma_{ij}(E)= e^2\tau(E)\int\frac{d^2k}{(2\pi)^2}\delta[E-E(k)]v^{i}(k)v^{j}(k),
\end{equation}
where $i,j=x,y$. For isotropic system $v_x^2=v_y^2=(1/2)(v_x^2+v_y^2)=(1/2)v^2$. In our case,
\begin{equation}
v^2=\frac{v_{F}^2}{2}\Big[1-\Big(\frac{\Delta_h}{E}\Big)^2\Big].
\end{equation}
Using these in Eq.[\ref{boltz}], the energy dependent conductivity becomes as
\begin{equation}\label{con}
\sigma(E)= e^2\tau(E)\frac{E}{\pi\hbar^2}
\Big[1- \Big(\frac{\Delta_h}{E}\Big)^2 \Big].
\end{equation}

Assuming the energy dependent scattering time to be 
$\tau=\tau_0 (E/E_{_F})^{m}$, where $m$ is a constant depending on the
scattering mechanism, $E_{F}=\sqrt{E_{F0}^2+\Delta_h^2}$
is the Fermi energy with $E_{F0}=\hbar v_{F}k_{F}^{0}$. 
Here, Fermi vector $k_{F}^{0}=\sqrt{2\pi n_e}$.
Substituting Eq. (\ref{con}) into Eq. (\ref{tp}),   
the diffusion thermopower is obtained as   
\begin{equation}
S= -L_0 \frac{e T}{E_{F0}} \Big[(m+1) +2\Big(\frac{\Delta_h}{E_{F0}}\Big)^2\Big]
/\sqrt{1+\Big(\frac{\Delta_h}{E_{F0}}\Big)^2}.
\end{equation}

We plot thermopower versus hybridization for different carrier density in 
the upper panel of Fig. [1]. 
It shows that thermopower increases with increasing hybridization
for a particular carrier density. But for higher carrier density,
this rate of enhancement with hybridization becomes very slow. 

Thermal conductivity can be directly obtained from Wiedemann-Franz law
given in Eq. (\ref{ther}) where the electrical conductivity $\sigma (E_{_F})$  
is given as
\begin{equation}
\sigma= \sigma_0/\sqrt{1+\Big(\frac{\Delta_h}{E_{F0}}\Big)^2}.
\end{equation}
Here, $ \sigma_0 =e^2\tau_{0}E_{F0}/(\pi\hbar^2)$ is the Drude conductivity without 
the hybridization constant for the Dirac system. Thermal conductivity is plotted in 
the the lower panel of Fig. [1]. 
We note that the thermal conductivity is diminished 
with increasing hybridization. However, unlike the case
of thermopower, here thermal conductivity increases with
carrier density.
\begin{figure}
\begin{center}
\includegraphics[width=90mm]{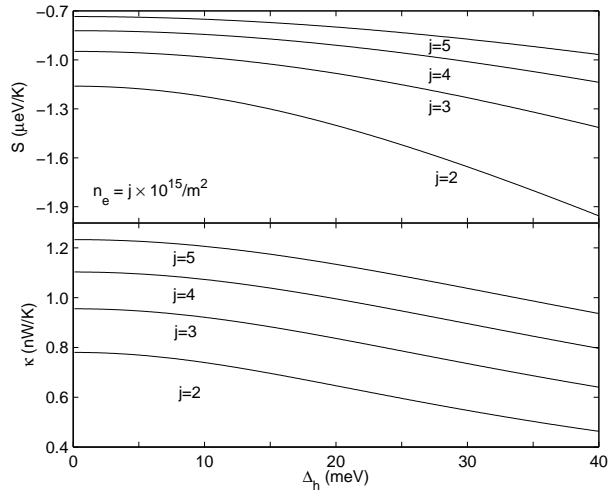}
\caption{Plots of the thermopower versus hybridization constant for $m=1$ and for
various carrier density.}
\end{center}
\end{figure}

\subsection{Non-zero magnetic field case}

In presence of magnetic field, the classical approach
can not explain the phenomenon depend on energy quantization. 
In this sub-section we follow quantum mechanical approach,
based on linear response theory, to study thermal transport coefficients.
Thermoelectric coefficients for two-dimensional electron system 
in presence of magnetic field were derived by modifying the Kubo formula in Ref. \cite{streda,oji}. 
These phenomenological transport coefficients are
\begin{equation}
\sigma_{\mu\nu} = {\cal L}_{\mu\nu}^{(0)}
\end{equation}
\begin{equation}
S_{\mu\nu} = \frac{1}{eT}[({\cal L}^{(0)})^{-1}{\cal L}^{(1)}]_{\mu\nu}
\end{equation}
\begin{equation}\label{cond1}
\kappa_{\mu\nu} = \frac{1}{e^2T}[{\cal L}_{\mu\nu}^{(2)}- 
eT({\cal L}^{(1)}S)_{\mu\nu}],
\end{equation}
where
\begin{equation}\label{Lmunu}
{\cal L}_{\mu\nu}^{(r)}=\int dE  \Big[-\frac{\partial f(E)}{\partial E}\Big]
(E-\eta)^{r}  \sigma_{\mu\nu}(E).
\end{equation}
Here, $\mu,\nu=x,y$. Also, $\sigma_{\mu \nu}(E)$,
$S_{\mu \nu} $ and $\kappa_{\mu \nu}$
are the zero-temperature energy-dependent conductivity, thermopower and
thermal conductivity tensors, respectively.

Generally, diffusive and collisional mechanism play major role
in electron conduction. The quantized energy spectrum of electrons
results itself through Shubnikov-de Hass oscillation by collisional mechanism.
In our case, electron transport is mainly due to the collisional
instead of diffusive. The zero drift velocity of electron
do not allow to have diffusive contribution. In presence of
temperature gradient, thermal transport coefficients can be expressed as
 $ {\cal L}_{xx}^{(r)}={\cal L}_{xx}^{(r){\rm col}} = 
{\cal L}_{yy}^{(r){\rm col}}$ and
$ {\cal L}_{yy}^{(r)}={\cal L}_{yy}^{(r){\rm dif}} + 
{\cal L}_{yy}^{(r){\rm col}}={\cal L}_{yy}^{(r){\rm col}}$.
In Ref. \cite{tahir2}, the exact form of the finite temperature collisional conductivity 
has been calculated for the screened impurity potential
$ U({\bf k}) = 2 \pi e^2/(\epsilon \sqrt{k^2 + k_s^2})\simeq 2\pi e^2/(\epsilon k_s) = U_0$ 
under the limit of small $|{\bf k}| \ll k_s $ with  $k_s$ and $ \epsilon $ being the inverse screening length 
and dielectric constant of the material, respectively. In this limit, one can 
use $\tau_0^2 \approx \pi l^2\hbar^2/N_IU_0^2$ with
$\tau_0$ is the relaxation time, 
$U_0$ is the strength of the screened impurity potential and 
$N_I$ is the two-dimensional impurity density.
The exact form of the finite temperature conductivity\cite{tahir2} 
can be reduced to the zero-temperature energy-dependent electrical conductivity as
\begin{equation} \label{exact}
\sigma_{xx}(E) = \frac{e^2}{h} 
\frac{ N_I U_0^2}{\pi \Gamma_0 l^2}\sum_{\tau_z}I^{\tau_z}_{n},
\end{equation}
where $I^{\tau_z}_{n} = [n\{1+\cos^2(\theta_{\tau_z})\}-\cos(\theta_{\tau_z})] $.
Here we have used $-\partial f/\partial E=\delta [E-E_n^{\tau_z}]$.
Using Eq. (\ref{Lmunu}), 
the finite temperature diagonal (${\cal L}_{xx}^{(r)} $) and off-diagonal 
coefficients  ($ {\cal L}_{yx}^{(r)}$) can be written as
\begin{equation}
{\cal L}_{xx}^{(r)} = 
\frac{e^2}{h}\frac{ N_I U_0^2}{\pi \Gamma_0 l^2}
\sum_{n,\tau_z}I^{\tau_z}_{n}\Big[(E-\eta)^{r}
\Big(-\frac{\partial f(E)}{\partial E}\Big)\Big]_{E=E^{\tau_z}_n} 
\end{equation}
and
\begin{eqnarray}
{\cal L}_{yx}^{(r)} & = & \frac{e^2}{h}\frac{1}{2}
\sum_{n,\tau_z}\frac{\sin ^2\theta_{\tau_z}}{\Delta_n^2}
\int_{E_n,\tau_z}^{E_{n+1},\tau_z}(E-\eta)^{r}
\Big(-\frac{\partial f(E)}{\partial E}\Big)dE.\nonumber\\
\end{eqnarray}
Here, $\Delta_n=\sqrt{2n+(\frac{\Delta_{\tau_z}}{\hbar\omega_c})^2}-\sqrt{2(n+1)
+(\frac{\Delta_{\tau_z}}{\hbar\omega_c})^2}$.

\section{Numerical results and discussion}

In our numerical calculations, the following parameters are used:
carrier concentration $n_e=2 \times 10^{15}$ m${}^{-2}$, 
$g = 60$, $v_F=4\times10^5$ m s${}^{-1}$ and $T=0.7$ K. These numerical parameters
are consistent with Ref.\cite{kehe,tahir1,tahir2}.

In Fig. [2], the off-diagonal thermopower, $S_{xy}$, as a function of 
the inverse magnetic field for different values of the hybridization 
constant is shown.
Similarly, the thermal conductivity, $ \kappa_{xx} $, versus 
inverse magnetic field for different values of the hybridization 
constant is shown in Fig. [3]. Careful observation of
these two figures clearly show that both $S_{xy}$ and  $ \kappa_{xx} $
oscillate with the same frequency which does not depend on  the 
hybridization energy. The hybridization energy only
influences the phase of oscillations.

\begin{figure}[t]
\begin{center}\leavevmode
\includegraphics[width=98mm]{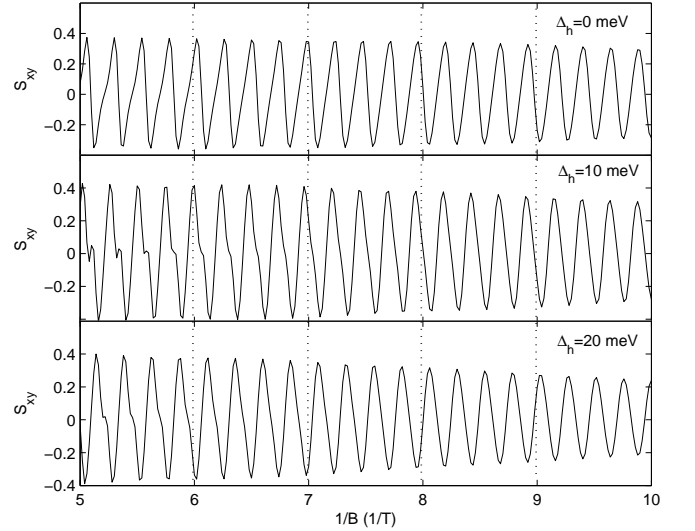}
\caption{Plots of the thermopower in units of $k_B/e$ versus 
inverse magnetic field for different values of the hybridization constant.}
\label{Fig1}
\end{center}
\end{figure}

\begin{figure}[t]
\begin{center}\leavevmode
\includegraphics[width=94mm,height=60mm]{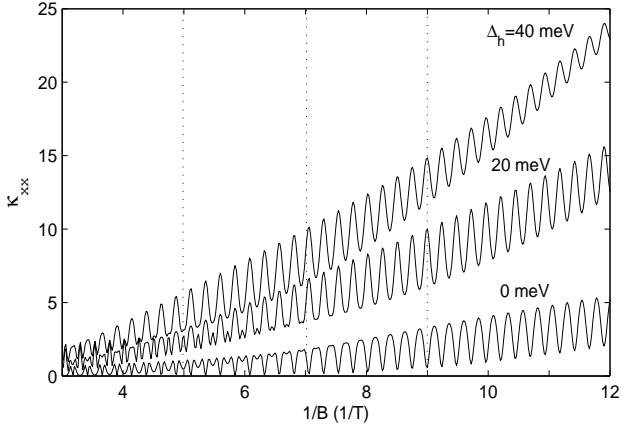}
\caption{Plots of the thermal conductivity in units of $k_B^2/h$ versus 
inverse magnetic field.}
\label{Fig1}
\end{center}
\end{figure}

To determine the frequency and the phase of the quantum oscillations in the thermoelectric 
coefficients, we shall derive analytical expressions of these coefficients.
The components of the thermopower are given by
\begin{equation} \label{Syy}
S_{xx} = S_{yy} = \frac{1}{eT}
\Big[ \frac{\sigma_{xx}}{S_0} {\cal L}_{xx}^{(1)}+
\frac{{\cal L}_{yx}^{(1)}}{\sigma_{yx}}\Big]
\end{equation}
and
\begin{equation} \label{Sxy}
S_{xy}= - S_{yx} = \frac{1}{eT}
\Big[\frac{\sigma_{xx}}{S_0} (-{\cal L}_{xy}^{(1)})+
\frac{{\cal L}_{xx}^{(1)}}{\sigma_{yx}} \Big].
\end{equation}
Here, $S_0=\sigma_{xx}\sigma_{yy}-\sigma_{xy}\sigma_{yx}$.
The dominating term in the above two equations is the last term. 
The analytical form of $\kappa_{xx}$ and $S_{xy}$ can be obtained directly by 
deriving analytical form of the phenomenological transport coefficients. 
The analytical form of density of states given in Eq. (\ref{dos_B}) allows us to obtain 
asymptotic expressions of $ S_{xy} $ and $ \kappa_{xx}$.
This is done by replacing the summation over discrete quantum 
numbers $n$ by the integration i.e; 
$ \sum_n \rightarrow 2\pi l^2 \int D_{\tau_z}(E)dE$,
then we get
\begin{equation}
{\cal L}_{xx}^{(1)} \simeq \frac{4\pi}{\beta}\frac{e^2}{h}\frac{\Gamma_0 E_F }
{(\hbar\omega_c)^2}\Omega_{D}G^{\prime}(x) \sum_{\tau_z}U_{\tau_z}
F_{\tau_z}\sin(2\pi F_{\tau_z})
\end{equation}
and
\begin{equation} \label{cond2}
{\cal L}_{xx}^{(2)} \simeq \frac{L_0 e^2T^2\sigma_0}{(\omega_c\tau_0)^2}
\sum_{\tau_z}U_{\tau_z}F_{\tau_z}[1 -3\Omega_{D} G^{\prime \prime}(x) 
\cos(2\pi F_{\tau_z})],
\end{equation}
where $F_{\tau_z}=(E_F^2-\Delta_{\tau_z}^2)/(\sqrt{2}\hbar\omega_c)^2$,
$U_{\tau_z}=[1+\cos^2(\bar{\theta}_{\tau_z})]$, the impurity induced damping factor is
\begin{equation}
\Omega_{D}=\exp\Big\{-\Big(2\pi\frac{\Gamma_0 E_F }{\hbar^2\omega_c^2}\Big)^2\Big\}
\end{equation}
and the temperature dependent damping factor is the derivative of 
the function $G(x) $ with $ G(x)=x/\sinh(x)$.
Here, $x=T/T_c$ with $T_c=(\hbar\omega_c)^2/(2\pi^2k_{_B}E_{F})$
is the critical temperature which depends on strength of hybridization
through Fermi energy.
Note that $G(x)$ is the temperature dependent damping factor for the electrical 
conductivity tensor.

The off-diagonal thermopower $S_{xy} $ and the diagonal thermal conductivity 
$\kappa_{xx}$ are obtained as given by
\begin{eqnarray}\label{ana_s}
S_{xy} & \backsimeq & \frac{ k_{_B}}{e} \frac{1}{k^0_Fl}\frac{16\pi}{\omega_c\tau_0} 
\Big[1+\Big(\frac{\Delta_h}{E_{F0}}\Big)^2\Big]^{1/2}\Omega_{D} G^{\prime}(x)\nonumber\\
& \times & \sum_{\tau_z}\frac{U_{\tau_z}}{\sin^2(\bar{\theta}_{\tau_z})}
F_{\tau_z}\sin\Big(2\pi f/B - \tau_z \phi \Big)
\end{eqnarray}
and
\begin{eqnarray} \label{diagok}
\kappa_{xx} &\simeq& L_0\frac{\sigma_0T}{(\omega_c\tau_0)^2}\sum_{\tau_z}
U_{\tau_z}F_{\tau_z} \nonumber \\ 
& \times & \Big[1 - 6\Omega_D G^{\prime \prime}(x)
\cos\Big(2\pi f/B - \tau_z \phi \Big) \Big],
\end{eqnarray}
where the frequency $f$ is given as
\begin{equation}
f=\frac{1}{2e\hbar v_{F}^2}(E_{F0}^2-\Delta_z^2)
\end{equation}
and the phase factor $\phi =  \pi g \mu_B \Delta_h/(e \hbar v_{F}^2)$.
Therefore, the thermopower and the thermal conductivity oscillate with the
same frequency $f$ which is independent of $\Delta_h$, which can be shown from numerical result also. 
The oscillation frequency is strongly reduced by the Zeeman energy $\Delta_z$.
On the other hand, the phase factor $ \phi $ is related to the product 
of the Lande $g$-factor and $\Delta_h$ and it vanishes if either of them is zero. 
Although the frequency and the phase of $ S_{xx} $ and $ \kappa_{xx}$ are the same but
the damping factor and amplitude are different.

Now we compare the numerical and analytical results of $s_{xy} $ and
$\kappa_{xx} $ in Fig. [4].
For better visualization, we have taken weak Landau level
broadening $\Gamma_0=0.01$ meV for $S_{xy}$ and $\kappa_{xx}$.
The analytical results, in particular the frequency $f$, match very well 
with the numerical results. 
We must mention here that for different values of 
$\Gamma_0$ analytical results may differ with numerical in the amplitude but 
the frequency and phase are always in good agreement.

\begin{figure}[t]
\begin{center}\leavevmode
\includegraphics[width=98mm]{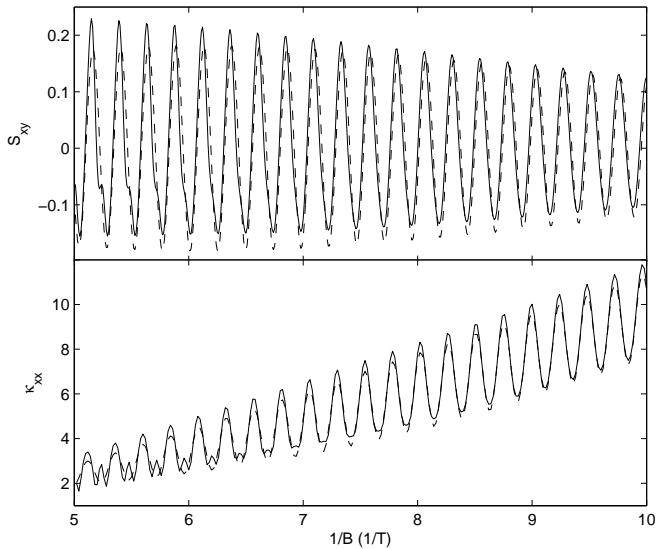}
\caption{Plots of the numerical and analytical results of the thermoelectric
coefficients versus inverse magnetic field for $\Delta_h=20$ meV. Solid and dashed lines 
stand for numerical and analytical results, respectively.}
\label{Fig1}
\end{center}
\end{figure}

It is interesting to see from the analytical expressions of the 
thermopower and the thermal conductivity that these transport coefficients 
possess weak periodic oscillation with the hybridization  energy for a 
given magnetic field. These oscillation is shown in figure 5..
The frequency and phase factor of these oscillations for a fixed $B$ are 
$ \nu = \phi/(2\pi) = g \mu_B /(2e \hbar v_{F}^2)$ and
$ \Phi = 2\pi f/B$, respectively. 

\begin{figure}[t]
\begin{center}\leavevmode
\includegraphics[width=90mm]{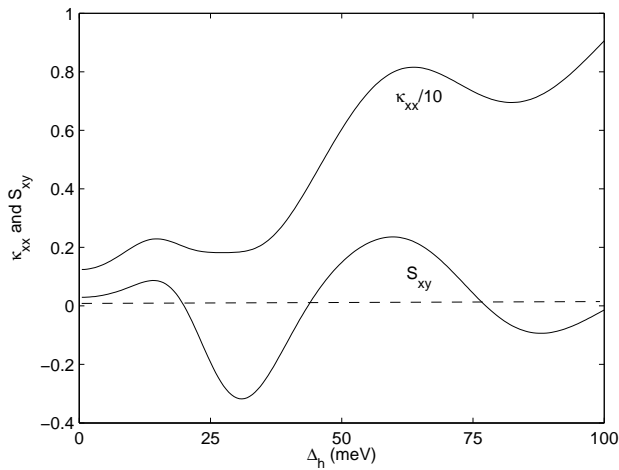}
\caption{Plots of the thermopower and thermal conductivity versus the 
hybridization constant $(\Delta_h)$. Here, $B=0.2$ T is taken.}
\label{Fig1}
\end{center}
\end{figure}


In presence of the magnetic field, these approximate analytical 
expression of the thermopower and thermal conductivity
can be also used for monolayer graphene by putting $\Delta_h=0$. 
There are several experimental results \cite{yuri,lau} on thermoelectric properties 
of a graphene monolayer but analytical expressions are not available in
the literature.
The approximate analytical expression of the thermopower and thermal conductivity
can also be used for a monolayer graphene by setting $\Delta_h=0$.



\section{conclusion}
We have presented a theoretical study on the 
thermoelectric coefficients of  ultra-thin topological insulators in
presence/absence of the magnetic field.
In absence of the magnetic field, the thermopower and the thermal conductivity
are modified due to the hybridization between top and bottom surface states.
The thermopower is enhanced and the thermal conductivity is diminished due to
the hybridization. 
The quantum oscillations in the thermopower and 
the thermal conductivity for different values of the hybridization constant
are also studied numerically. 
In addition to the numerical results, we obtained the analytical expressions 
the thermopower $(S_{xy})$ and the thermal conductivity ($\kappa_{xx}$). 
The analytical results match very well with the numerical results.
We have also provided analytical expressions of the oscillation frequency
and phase. The oscillation frequency is the same for both the thermopower and 
thermal conductivity. It is independent of the hybridization constant but 
strongly suppressed by the Zeeman energy.
On the other hand,
the hybridization constant plays a very significant role in the phase as well as in
the amplitude of the oscillations. From the analytical results,
critical temperature $(T_c)$ is found to be reduced with increasing 
hybridization constant. Thermoelectric coefficients also show
a very low-frequency oscillation with the hybridization constant for a given
magnetic field. 
Moreover, our analytical expressions of the thermopower and
thermal conductivity are also applicable for a graphene monolayer
by setting $ \Delta_h=0$.

\section{Acknowledgement}
This work is financially supported by the CSIR, Govt.of India under the grant
CSIR-SRF-09/092(0687) 2009/EMR F-O746.

\begin{appendix}

\section{}
Derivation of asymptotic analytical expression of density of
states of a two dimensional electronic system in presence of impurity
can be done by calculating self-energy \cite{ando,gerhartz} which is given as
\begin{equation}\label{sum}
\Sigma^-(E)=\Gamma_0^2\sum_n \frac{1}{E-E_{n}^{\tau_z}-\Sigma^-(E)}.
\end{equation}
Imaginary part of the self-energy is related to density of states as
$ D(E)= \Im \left[\frac{\Sigma^-(E)}{\pi^2 l^2 \Gamma_0^2}\right] $.

By using residue theorem, we calculate the summation in Eq.(\ref{sum}), which give
$ \Sigma^-(E)\simeq\frac{2\pi\Gamma_0^2 E}{(\hbar\omega_c)^2}\cot(\pi n_0)$, 
where $n_0$ is the pole  and given as
\begin{equation}
 n_0 = \frac{1}{2(\hbar\omega_c)^2}\Big[\{E-\Sigma^{-}(E)\}^2-(\Delta_z+\tau_z\Delta_h)^2\Big].
\end{equation}

By writing the self-energy as the sum of real and imaginary part, it can be further simplified as
\begin{equation}
\Delta + i \frac{\Gamma}{2} \simeq \frac{2\pi\Gamma_0^2 E}{(\hbar\omega_c)^2} 
\cot\Big[\frac{(u-iv)}{2} \Big]
\end{equation}
Here,
\begin{equation}
 u = \frac{\pi}{(\hbar\omega_c)^2}[E^2-(\Delta_z+\tau_z\Delta_h)^2]
\end{equation}
 and
$ v = \pi\Gamma E/(\hbar\omega_c)^2 $.
The imaginary part is
$ \frac{\Gamma}{2} = \frac{2\pi\Gamma_0^2 E}{(\hbar\omega_c)^2}\
\frac{\sinh v}{\cosh v-\cos u} $.
Now, this can be re-written by using the following standard relation as
\begin{equation}
\frac{\sinh v}{\cosh v-\cos u} = 1 + 2 \sum_{s=1}^{\infty} e^{-s v} \cos (su).
\end{equation}
Here, the  most dominant term is for $s=1 $ only. 
We can write
\begin{equation}
\frac{\Gamma}{2}  =  \frac{2\pi\Gamma_0^2 E}{(\hbar\omega_c)^2}
\Big [1 + 2 \sum_{s=1}^{\infty} e^{-s\pi \Gamma E/(\hbar \omega_c)^2} \cos(u) \Big].
\end{equation}
In the limit of $\pi\Gamma \gg \hbar\omega_c $, after first iteration, we have
$ \Gamma/2=2 \pi\Gamma_0^2 E/(\hbar\omega_c)^2$.
Substituting it in the earlier expression, we get
\begin{eqnarray}
\frac{\Gamma}{2} & = & \frac{2\pi\Gamma_0^2 E}{(\hbar\omega_c)^2}
\Big [1+2 \sum_{s=1}^{\infty}\exp\Big\{-s\Big(\frac{2\pi\Gamma_0 E}{\hbar^2\omega_c^2}\Big)^2\Big\} 
\nonumber\\
& & 
\cos\Big\{s\pi\Big(E^2-\Delta_{\tau_z}^2\Big)/(\hbar\omega_c)^2\Big\} \Big].
\end{eqnarray}
Here, $\Delta_{\tau_z}=\Delta_z+\tau_z\Delta_h$

Finally, the density of states for two branches can be obtained as
\begin{eqnarray}
D_{\tau_z}(E)&=&\frac{D_0(E)}{2}\Big [1+2 \sum_{s=1}^{\infty}\exp\Big\{-s\Big(\frac{2\pi\Gamma_0 E}
{\hbar^2\omega_c^2}\Big)^2\Big\} 
\nonumber\\
& & 
\cos\Big\{s\pi\Big(E^2-\Delta_{\tau_z}^2\Big)/(\hbar\omega_c)^2\Big\} \Big].
\end{eqnarray}
\end{appendix}

\end{document}